\documentclass{article}

\usepackage{microtype}
\usepackage{graphicx}
\usepackage{subfigure}
\usepackage{booktabs} 

\usepackage[hyphens]{url}
\usepackage{hyperref}

\usepackage[accepted]{icml2024}

\usepackage{amsmath}
\usepackage{amssymb}
\usepackage{mathtools}
\usepackage{amsthm}

\usepackage[capitalize,noabbrev]{cleveref}

\theoremstyle{plain}

\theoremstyle{definition}

\theoremstyle{remark}

\usepackage[disable,textsize=tiny]{todonotes}

\icmltitlerunning{Training Foundation Models as Data Compression}

\begin{document}

\twocolumn[
\icmltitle{Training Foundation Models as Data Compression: \\On Information, Model Weights and Copyright Law}




\begin{icmlauthorlist}
\icmlauthor{Giorgio Franceschelli}{unibo}
\icmlauthor{Claudia Cevenini}{unibo}
\icmlauthor{Mirco Musolesi}{ucl,unibo}
\end{icmlauthorlist}

\icmlaffiliation{unibo}{Department of Computer Science and Engineering, Alma Mater Studiorum - Università di Bologna, Bologna, Italy}
\icmlaffiliation{ucl}{Department of Computer Science, University College London, London, United Kingdom}

\icmlcorrespondingauthor{Giorgio Franceschelli}{giorgio.franceschelli@unibo.it}

\icmlkeywords{Machine Learning, ICML}

\vskip 0.3in
]



\printAffiliationsAndNotice{}  

\begin{abstract}
The training process of foundation models as for other classes of deep learning systems is based on minimizing the reconstruction error over a training set. For this reason, they are susceptible to the memorization and subsequent reproduction of training samples. In this paper, we introduce a \textit{training-as-compressing} perspective, wherein the model’s weights embody a compressed representation of the training data. From a copyright standpoint, this point of view implies that the weights can be considered a reproduction or, more likely, a derivative work of a potentially protected set of works. We investigate the technical and legal challenges that emerge from this framing of the copyright of outputs generated by foundation models, including their implications for practitioners and researchers. We demonstrate that adopting an \textit{information-centric approach} to the problem presents a promising pathway for tackling these emerging complex legal issues.
\end{abstract}

\section{Introduction}

Besides curiosity and excitement, the current generative AI wave raises various philosophical and practical questions \cite{shanahan2024talking,weidinger2022taxonomy}. Among them, a relevant issue is how current copyright laws can be applied to generative AI \cite{lee2024talkin}. For instance, can AI-generated outputs be copyrighted? Is it lawful to train models on protected works? Furthermore, would it be possible to protect the model itself in some manner?

To better understand how such models work and how they can be interpreted under current laws, we believe that a viable solution is to link generative deep learning with information theory~\cite{cover1999elements}. In fact, generative models trained with self-supervised learning, i.e., by maximizing the likelihood of training data, as commonly done for foundation models~\cite{bommasani2021opportunities}, such as large language models (LLMs)~\cite{team2023gemini,meta2024introducing} and diffusion models~\cite{rombach2022high}, can be seen as a form of \textit{(lossy or lossless) compression}~\cite{mackay2003information}. From this perspective, the training algorithm plays the role of the compression algorithm; the inference (feed-forward) algorithm is the de-compression algorithm (with the input passed to the model working as a decoding key); and the model's weights represent the compressed version of the training set.

\begin{figure}[t]
    \centering
    \includegraphics[width=.48\textwidth]{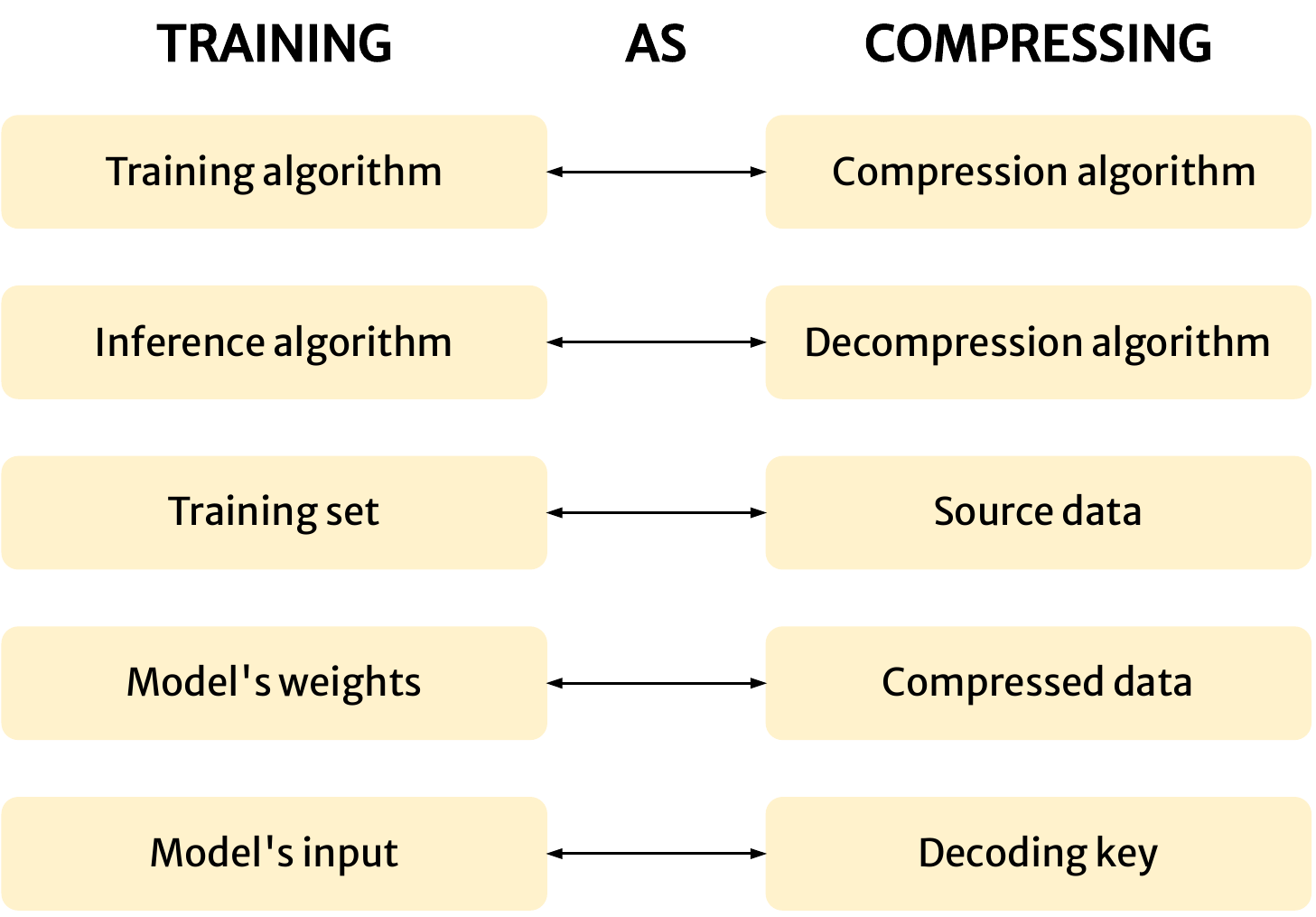}
    \caption{The \textit{training-as-compressing} perspective. The training set is compressed into the weights via a training algorithm; the source data can be retrieved using the appropriate model's input.}
    \label{fig:t_as_c}
\end{figure}

Deletang et al. \cite{deletang2024language} discuss how a language model can implement a lossless compression process offline, i.e., through a fixed set of model parameters derived from training. We move a step further and, following \cite{cooper2024files}, claim the self-supervised learning used to train foundation models to be a lossy or lossless compression process, during which all training data are compressed into the model's weights at different levels of lossiness.

This is demonstrated by the fact that the model can reproduce certain portions of training samples \cite{carlini2023quantifying}. We suggest that the training optimizes the model’s weights to be the best possible compressed version of the training set, or more accurately, batches of it at a time. The analogies at the basis of the proposed \textit{training-as-compressing} perspective are summarized in Figure \ref{fig:t_as_c}.
Building on this intuition, we envision the model's weights as either a reproduction or a derivative work of training data. This new interpretation opens up a series of practical consequences that can be relevant from the copyright perspective. First, it provides a legal framework for understanding generative models which relates them to the training data.
Then, it creates a direct link between the training data and the model outputs only made of steps requiring specific exceptions or authors' authorization, thus allowing for potential requests for compensation. Our legal discussion primarily centers on EU laws; however, many of the considerations discussed can be broadly applied to other countries that adhere to the Berne Convention.

This article is structured as follows. In Section \ref{sec:information}, we introduce self-supervised learning, i.e., the training scheme behind foundation models, as a form of data compression. Then, we discuss how our \textit{training-as-compressing} perspective allows for a specific understanding of the model's weights under copyright law (Section \ref{sec:copyright}). Finally, we discuss the legal implications of the framing presented in this work (Section \ref{sec:implications}).

\section{Training-as-Compressing and Information} \label{sec:information}

The propensity of foundation models to memorize and subsequently replicate training data is a topic that has received considerable attention in scholarly literature, as evidenced by works such as \cite{carlini2023extracting}. In general, it is very hard to decompress every possible training sample perfectly and in its entirety, i.e., without any loss of information. Nonetheless, it has been shown that training samples can be retrieved \cite{carlini2021extracting}, and more advanced techniques might lead to an even higher degree of reconstruction \cite{cooper2024files}. The following thought experiment may help us understand this matter better.

\begin{figure*}[t]
    \centering
    \includegraphics[width=.95\textwidth]{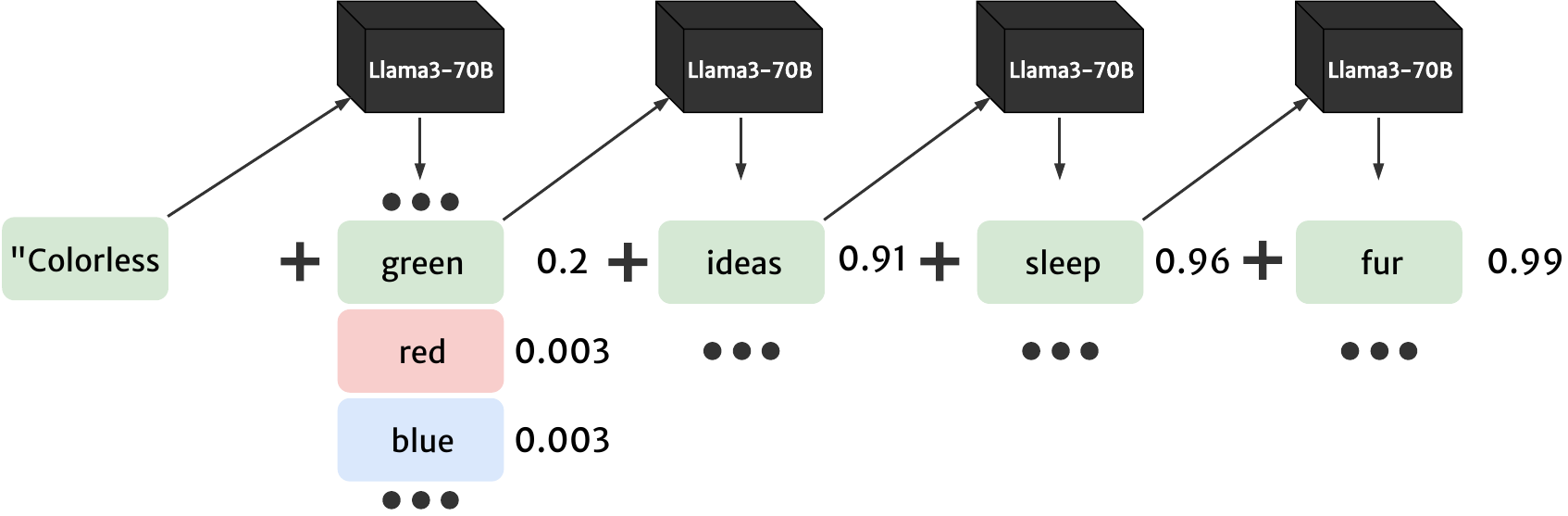}
    \caption{A thought experiment to confirm that LLMs memorize training data: even if the sentence is semantically nonsensical, the model assigns high probability to its tokens just because it occurred in its training set.}
    \label{fig:chomsky}
\end{figure*}

In 1957, Noam Chomsky introduced the famous sentence ``Colorless green ideas sleep furiously'' to demonstrate the distinction between syntax and semantics: the sentence is grammatically well-formed but semantically nonsensical. If a language model had learned the semantics of English, it should not generate a semantically nonsensical sentence, i.e., it should assign to semantically nonsensical words a small probability; if, on the contrary, such words are characterized by a large probability of being generated, then it is very likely that the model has memorized them. To test this, we use the same quote from Chomsky, and we check the probability of each subsequent word given the previous ones under an LLM (in our case, LLaMa3-70B \cite{meta2024introducing}). As depicted in Figure \ref{fig:chomsky}, the probability of \textit{` green'} given \textit{` ``Colorless'} is 0.2, while for other colors like \textit{` red'} or \textit{` blue'} is 0.003. For all the subsequent words, i.e., \textit{` ideas'}, \textit{` sleep'}, and \textit{` furiously'}, the probability is always greater than 0.9. 
The model has essentially memorized the quote into its weights, otherwise it would have never assigned such high probabilities to a semantically nonsensical sentence.

We know that an LLM memorizes at least parts of its training data. Can we also say it compresses them?
Consider again LLaMa3-70B \cite{meta2024introducing}. This model is pre-trained on more than 15 trillion tokens. Each token can have one out of $128000 \approx 2^{17}$ values, thus requiring at least 17 bits to be represented. This means the training data require at least $17 \cdot 15 \cdot 10^{18} = 255 \cdot 10^{18}$ bits to be memorized. However, the model has 70 billion weights and uses half-precision floating points (i.e., $16$ bits for each weight), thus requiring a total of $16 \cdot 70 \cdot 10^{15} \approx 1.1 \cdot 10^{18}$ bits. With smaller models such as LLaMa3-8B, the compression ratio is even more astonishing and may contribute to the lower degree of exact or near-exact reproduction of training samples \cite{carlini2023extracting}.

Indeed, a foundation model, such as a transformer-based LLM, consists of a neural network with weights $\mathbf{W}$. It models the conditional distribution $P(x_i | x_{i-k}, ..., x_{i-1}, t; \mathbf{W})$, where $x = x_1, ..., x_N$ is the tokenized input to be modeled (and also the output to be predicted, which is the reason of the \textit{self-}supervised learning), $k$ is the size of the context window, and $t$ is an additional input (such as a task specification). During training, the randomly initialized weights are iteratively adjusted through stochastic gradient descent (and its variants) as follows:
\begin{equation}
    \mathbf{W} \gets \mathbf{W} - \alpha \frac{1}{|\mathbb{X}|}\nabla_{\mathbf{W}}L(\mathbb{X}, \mathbf{W}),
\end{equation}
\noindent where $\alpha$ is the learning rate and $L(\mathbb{X}, \mathbf{W})$ is the loss function computed on a batch of training samples $\mathbb{X}$. In particular, the objective is to maximize the log-likelihood of training data, therefore the loss is defined as:
\begin{equation}
    L(\mathbb{X}, \mathbf{W}) = - \sum_{x,t \in \mathbb{X}} \sum_{i} \log P(x_i | x_{i-k}, ..., x_{i-1}, t; \mathbf{W}).
\end{equation}
In other words, the training phase aims to find the optimal values of the weights $\mathbf{W}$ such that given the input $t$ (i.e., the decoding key) the model can autoregressively reconstruct $x$ by only using the information stored into $\mathbf{W}$.

From an information theoretic perspective, such training data compression can be explained through the information bottleneck (IB) principle \cite{tishby1999information}. The IB principle applies when we aim to extract relevant information from an input variable $X \in \mathcal{X}$ about an output variable $Y \in \mathcal{Y}$. Given their joint distribution $p(X, Y)$, the relevant information is defined as the mutual information $I(X; Y)$. With $\hat{X}$ as the relevant part of $X$ with respect to $Y$, the IB method aims to find the optimal $\hat{X} \in \mathcal{\hat{X}}$, i.e., the one that minimizes $I(X;\hat{X})$ (obtaining the simplest possible statistics) while maximizing $\beta I(\hat{X};Y)$ (containing all the relevant information). Tishby and Zaslavsky \cite{tishby2015deep} argued that neural networks could be interpreted under the theoretical framework of the IB principle. Indeed, neural networks learn to extract efficient representations of the relevant features $\hat{X}$ of the input $X$ for predicting the output $Y$, given a finite sample of the joint distribution $p(X, Y)$. In the context of supervised learning, this means ignoring the irrelevant part of $X$ by only selecting the one needed to predict $Y$. However, in self-supervised learning $Y \approx X$. It follows that $\hat{X}$ is the relevant part of $X$ with respect to itself, so it is a \textit{compressed} version of $X$.

These considerations suggest a \textit{training-as-compressing} analogy, where the training algorithm plays the role of the compression algorithm; the inference (feed-forward) algorithm is the de-compression algorithm (with the input passed to the model working as a decoding key); and the model's weights represent the compressed version of the entire training set.
However, due to the stochasticity of the learning process and the fact that some data can occur multiple times in the training set, different training samples are compressed with a different degree of lossiness, i.e., some can be reconstructed (near-)exactly, while other reconstructions will be too lossy \cite{cooper2024files}.

In addition to being used for data generation \textit{as-is}, such a \textit{pre-trained} model commonly represents a starting point for additional training: it can be fine-tuned for supporting downstream tasks \cite{tunstall2022natural} or inducing desired behaviors, e.g., to align it with human preferences \cite{leike2018scalable}.
The discussion above can be extended to fine-tuned models as well. The only difference is that the source data would be both the additional training set and the pre-trained model's weights.

\section{Training-as-Compressing and Copyright Interpretation} \label{sec:copyright}

While the software code responsible for the training and inference of a generative model can fall under copyright law as a computer program, the algorithmic method is considered a mathematical model and thus not protected \cite{wto1994agreement}. However, the interpretation of the model's weights remains an open question.

Indeed, if the model's weights represent a compressed version of the training set, and copyright laws protect the training set, then the weights are also subject to them \cite{lee2024talkin}. Assuming that the training set is protected in some ways (we will discuss it later), the weights can thus be seen as either a) a lossy or lossless compressed copy or b) a compressed version of a derivative work.

Seeing the weights as a mere compressed copy of the training set (not different from a zipped file) is seducing since the weights are meant to contain all the information necessary to reconstruct the original samples given a certain input (i.e., the decoding key). However, the final result is usually lossy, and the common scenario is that what we obtain after decompression is similar, but not exactly equal, to the original work. If the differences are not substantial, then it can still be considered a copy; however, it can also lead to a non-negligible modification or transformation of the training data. This second option matches the definition of derivative works.

This opens up a different perspective: what the weights represent might not be the original training set, but a new, derivative work (substantially different from, but still based on, the original) \cite{lee2024talkin} whose creation happens concurrently with weights' learning and whose only existence is due to the weights themselves. Nonetheless, a derivative work must still satisfy the originality requirement to be protected by copyright. Whether or not the trainers' role in choosing data, algorithms, and parameters is sufficient for claiming authorship (and thus protection) of the model's weights is still an open question.

Until now, we have assumed that the training set is protected under copyright law. The whole training set can be protected as a database or a collective work, i.e., a collection of separate and independent works \cite{lee2024talkin}. However, the collective work must constitute an intellectual creation because of the selection and arrangement of its content; the same criteria also apply to databases. One of the current trends for training foundation models seems to go in the opposite direction. Although a certain degree of data pre-processing is always present, the apparent tendency, at least in the early days of foundation models, has been to collect as much data as possible, for example, from the Web. This approach threatens the requirement of making a careful and original selection or arrangement. 
Moreover, training sets for fine-tuning models on specific domains are more likely to be eligible for protection as collective works.
Still, this interpretation does not cover all foundation models' training sets.

On the other hand, single training samples are often protected under copyright law \cite{bandy2021addressing}. Even though the training goal aims to compress batches of samples at a time, thus potentially leading to a compression that is optimal for a subset of works when considered together but not when considered separately, the single works can still be decompressed from the resulting model, at least in principle. 
This suggests that the model's weights can be interpreted as containing either a copy or a derivative work of any of the independent training samples (depending on their degree of compression \cite{cooper2024files}), and not only of the training data as a whole.

\section{Implications} \label{sec:implications}

Interpreting the model's weights as a copy or a derivative work of protected works leads to two crucial implications.

First, it provides a legal framework to understand them, removing the veil of uncertainty surrounding this issue. Although asserting copyright protection for weights as a derivative work presents challenges due to the absence of valid authorship \cite{otero2021machine}, 
it is arguably possible to safeguard them as databases of weights (i.e., collections of independently retrievable floating-point numbers) through the \textit{sui generis right} (thus providing certain rights to those who have invested in the database constitution independently from its copyright protection) \cite{sousa2024ai}.
In other words, the \textit{sui generis right} can protect the investment; our copyright perspective can link the model's weights back to the training data, providing a new interpretation of one of the several issues concerning the generative-AI supply chain \cite{lee2024talkin}. The same considerations still hold in the case of a fine-tuned model. According to Lee et al. \cite{lee2024talkin}, this would be considered a derivative work of the pre-trained model (and also of the fine-tuning data). In other words, fine-tuning is nothing more than an additional step in the information processing chain. Again, the weights of the fine-tuned model would be eligible for the \textit{sui generis right}. However, whether it qualifies for protection as a derivative work remains an open question, and the determination of valid (human) authorship can vary on a case-by-case basis.

\begin{figure*}[t]
    \centering
    \includegraphics[width=.95\textwidth]{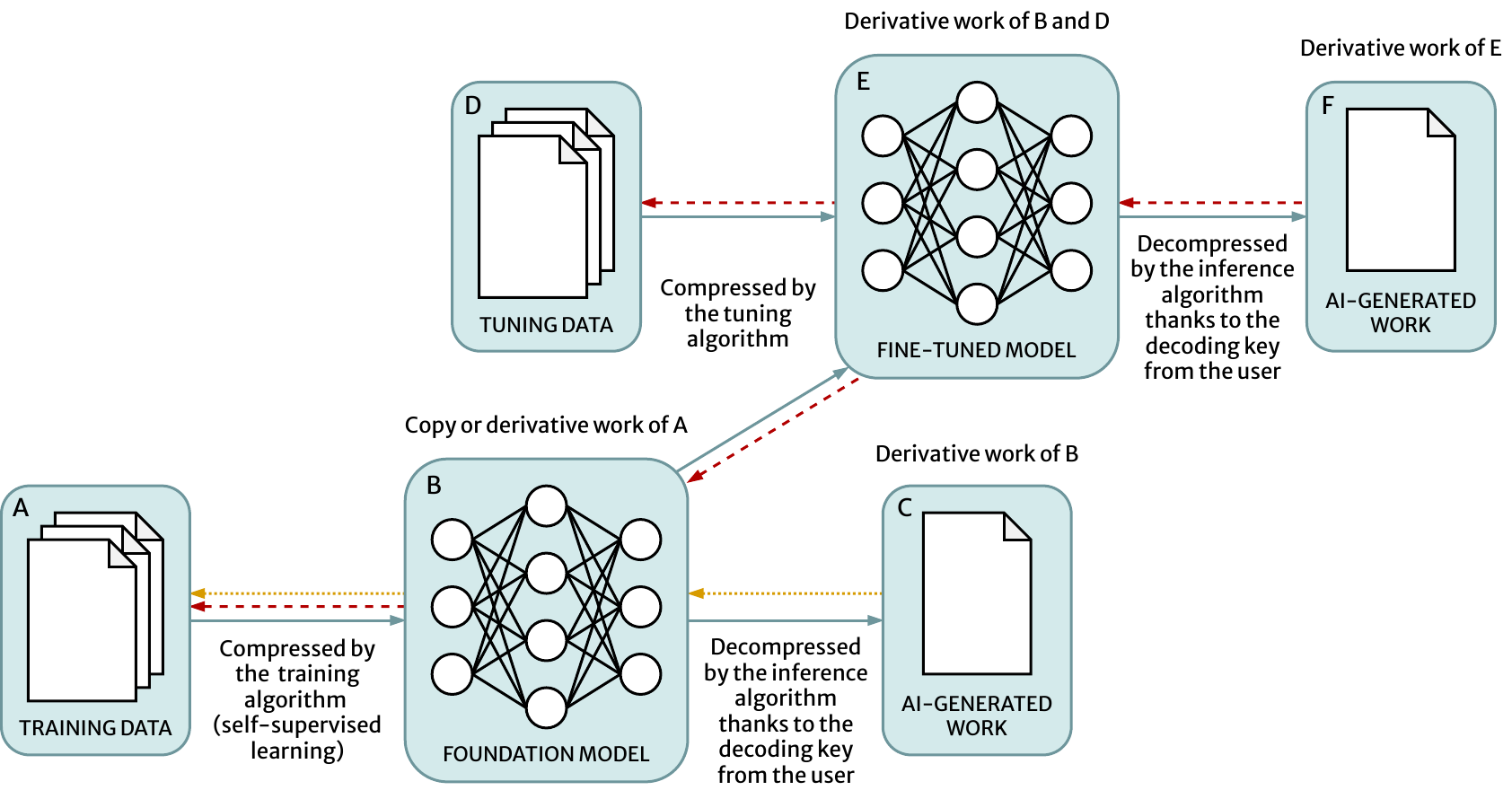}
    \caption{A schematic summary of the legal framework resulting from the \textit{training-as-compressing} perspective. The blue arrows connect potentially protected entities to their copies or derivative works: the foundation model is a copy or a derivative work of the training data; fine-tuning leads to a new derivative work of the foundation model and the tuning data; and an AI-generated work is a derivative work of either the foundation or the fine-tuned model. The yellow (dotted) and red (dashed) arrows directly link the AI-generated work back to training data and training and tuning data, respectively, only through steps requiring specific exceptions or authorization.}
    \label{fig:summary}
\end{figure*}

Second, this type of interpretation provides a potential framework for works generated by the model. Indeed, decompressing the information from the model can be seen as producing a derivative work of the weights, thus a derivative work of a copy of a protected work or a derivative work of a derivative work of (a copy of) a protected work. Either way, this link between the output and the training data can help enforce their copyrights. It is worth noting that the EU text and data mining (TDM) exceptions as well as other comparable rules \cite{fiilflynn2022legal} apply for TDM purposes, such as training the model, therefore to the case of the creation of a copy or derivative work (although TDM exceptions explicitly refer only to extraction and reproduction); however, they do not apply for further derivative works from the model. A similar consideration can also be drawn for the US fair use doctrine, which arguably applies to training a model on copyrighted data but is less likely when deployed to generate similar content that can threaten their market \cite{henderson2023foundation}.
The main consequence is that generating an output with the trained model would require authorization from the training set's rightsholders (or else the reproduction or adaptation right would be triggered), allowing for potential requests for compensation from the original authors.
In addition, generated works would need to respect the moral rights (e.g., the right to be identified as the author) of the owners of training data, even when their economic rights have expired. The fact that a new derivative work is protected by copyright is an entirely different issue that will depend on the human contribution (i.e., the input to the model), in particular, on its substantiality and its being the main contribution of the originality \cite{guadamuz2017androids,franceschelli2022copyright}. Crucially, these considerations also apply to synthetic data (i.e., AI-generated works) used as new training data for a different foundation model (e.g., \cite{shumailov2024ai}). Essentially, the chain of relationships between the model-generated data and the original protected training data would be longer but still involve steps that would trigger either the reproduction or adaptation rights.
The overall conceptual framework based on the \textit{training-as-compressing} perspective is summarized in Figure \ref{fig:summary}. 

\section{Related Work}

\subsection{Deep Learning and Information Theory}

Information theory is underpinning machine learning \cite{mackay2003information}.
The problem of training data compression has been analyzed from the point of view of the information bottleneck principle (IB) \cite{tishby1999information}, which provides a methodology for extracting a quantitative measurement of ``meaningful'' and ``relevant'' information.
In particular, IB can be used to derive an optimal theoretical limit for training neural networks in terms of trade-off between compression and prediction \cite{tishby2015deep}. From this perspective, stochastic gradient descent can be seen as composed of two distinct phases: a ``fitting'' phase in which the mutual information between the weights and the target output increases and a ``compression'' phase in which the mutual information between the weights and the input decreases \cite{shwartzziv2017opening}. However, the latter phenomenon might not be an intrinsic feature of deep neural networks, but only the result of the presence of nonlinearities in the networks \cite{michael2018on}. An important recent paper in this context is \cite{shwartzziv2024compress}, in which Schwartz-Ziv and LeCun discuss IB in relation to self-supervised learning, especially in the context of multimodal learning.
Finally, in a concurrent work, Cooper and Grimmelmann \cite{cooper2024files}  discuss memorization and compression in foundation models, arguing that some training data are losslessly memorized and others are compressed. In this work, we analyze the problem from an information-theoretic point of view, with a focus on the problem of fine-tuning and self-supervised training as well.

\subsection{Generative AI and Copyright}

There has been a long-standing interest in copyright and generative AI \cite{butler1982computer,samuelson1986allocating}.
Different legal issues are at play when considering the entire generative AI supply chain \cite{lee2024talkin}. Whether training neural networks on protected data is lawful has been highly debated across different national legislations \cite{lemley2021fair}, e.g., U.S. fair use \cite{henderson2023foundation}, EU text and data mining exceptions \cite{sobel2020taxonomy}, and others \cite{guadamuz2024scanner,samuelson2021copyright}. The other main debate has focused on whether a machine-generated work is protected by copyright \cite{craig2019death,grimmelman2016thing}, and who might own its copyright in the future \cite{franceschelli2022copyright}. However, other issues have also been considered, such as whether the model output can infringe the reproduction right \cite{vyas2023provable} or how the trained model can be protected \cite{otero2021machine}.
Finally, Schwarzschild et al. \cite{schwarzschild2024rethinking} recently proposed a new metric for memorization based on the compressibility (i.e., the shortness) of the prompt eliciting the reconstruction of a training work at inference time, and suggested its applications in the legal context. While similar in spirit, our notion of compression refers to the process of memorization at training time, and our legal considerations are complementary to theirs.

\section{Conclusion}

Motivated by the phenomenon of data memorization in foundation models, we have proposed a perspective we termed as \textit{training-as-compressing} for examining copyright issues. Starting from a thought experiment and a theoretical discussion, we have suggested interpreting self-supervised learning as data compressing and the model's weights as a compressed version of the entire training set. From a copyright perspective, this framing has led us to consider the weights as a reproduction or a derivative work of training data, which usually contain protected works. 
Finally, we have discussed the potential repercussions of this framework and the emergence of a direct relationship between the AI-generated outputs and the protected training data.

The analysis conducted in this paper predominantly raises additional questions rather than providing unequivocal solutions.
We believe that a multi-disciplinary analysis of these problems and implications from the point of view of information theory is of fundamental importance for practitioners and researchers from both the technological and legal points of view. Our research agenda also includes a rigorous formalization of the problem as a basis for rigorous legal analysis of this complex yet fascinating area.

\section*{Acknowledgments}

The participation and presentation at GenLaw'24 was supported by the ISA Doctoral Prize (ISA DP), offered by Istituto di Studi Avanzati, Alma Mater Studiorum Università di Bologna.

\bibliography{biblio}

\begin{thebibliography}{40}
\providecommand{\natexlab}[1]{#1}
\providecommand{\url}[1]{\texttt{#1}}
\expandafter\ifx\csname urlstyle\endcsname\relax
  \providecommand{\doi}[1]{doi: #1}\else
  \providecommand{\doi}{doi: \begingroup \urlstyle{rm}\Url}\fi

\bibitem[Bandy \& Vincent(2021)Bandy and Vincent]{bandy2021addressing}
Bandy, J. and Vincent, N.
\newblock Addressing ''documentation debt'' in machine learning: A retrospective datasheet for bookcorpus.
\newblock In \emph{{Proc. of the 35th Conference on Neural Information Processing Systems Datasets and Benchmarks Track (Round 1)}}, 2021.

\bibitem[Bommasani et~al.(2021)Bommasani, Hudson, Adeli, Altman, Arora, von Arx, Bernstein, Bohg, Bosselut, Brunskill, et~al.]{bommasani2021opportunities}
Bommasani, R., Hudson, D.~A., Adeli, E., Altman, R., Arora, S., von Arx, S., Bernstein, M.~S., Bohg, J., Bosselut, A., Brunskill, E., et~al.
\newblock On the opportunities and risks of foundation models.
\newblock \emph{arXiv preprint arXiv:2108.07258}, 2021.

\bibitem[Butler(1982)]{butler1982computer}
Butler, T.~L.
\newblock Can a computer be an author? copyright aspects of artificial intelligence.
\newblock \emph{Hastings Communications and Entertainment Law Journal}, 4\penalty0 (4):\penalty0 707, 1982.

\bibitem[Carlini et~al.(2021)Carlini, Tram{\`e}r, Wallace, Jagielski, Herbert-Voss, Lee, Roberts, Brown, Song, Erlingsson, Oprea, and Raffel]{carlini2021extracting}
Carlini, N., Tram{\`e}r, F., Wallace, E., Jagielski, M., Herbert-Voss, A., Lee, K., Roberts, A., Brown, T., Song, D., Erlingsson, {\'U}., Oprea, A., and Raffel, C.
\newblock Extracting training data from large language models.
\newblock In \emph{{Proc. of the 30th USENIX Security Symposium (USENIX Security 21)}}, 2021.

\bibitem[Carlini et~al.(2023{\natexlab{a}})Carlini, Hayes, Nasr, Jagielski, Sehwag, Tram\`{e}r, Balle, Ippolito, and Wallace]{carlini2023extracting}
Carlini, N., Hayes, J., Nasr, M., Jagielski, M., Sehwag, V., Tram\`{e}r, F., Balle, B., Ippolito, D., and Wallace, E.
\newblock Extracting training data from diffusion models.
\newblock In \emph{{Proc. of the 32nd USENIX Conference on Security Symposium (SEC'23)}}, 2023{\natexlab{a}}.

\bibitem[Carlini et~al.(2023{\natexlab{b}})Carlini, Ippolito, Jagielski, Lee, Tramer, and Zhang]{carlini2023quantifying}
Carlini, N., Ippolito, D., Jagielski, M., Lee, K., Tramer, F., and Zhang, C.
\newblock Quantifying memorization across neural language models.
\newblock In \emph{{Proc. of the 11th International Conference on Learning Representations (ICLR'23)}}, 2023{\natexlab{b}}.

\bibitem[Cooper \& Grimmelmann(2024)Cooper and Grimmelmann]{cooper2024files}
Cooper, A.~F. and Grimmelmann, J.
\newblock The files are in the computer: Copyright, memorization, and generative ai, 2024.
\newblock Forthcoming, \textit{Chicago-Kent Law Review}.

\bibitem[Cover(1999)]{cover1999elements}
Cover, T.~M.
\newblock \emph{Elements of Information Theory}.
\newblock John Wiley \& Sons, 1999.

\bibitem[Craig \& Kerr(2019)Craig and Kerr]{craig2019death}
Craig, C.~J. and Kerr, I.~R.
\newblock The death of the {AI} author.
\newblock \emph{Ottawa Law Review}, 52\penalty0 (1):\penalty0 31--86, 2019.

\bibitem[Deletang et~al.(2024)Deletang, Ruoss, Duquenne, Catt, Genewein, Mattern, Grau-Moya, Wenliang, Aitchison, Orseau, Hutter, and Veness]{deletang2024language}
Deletang, G., Ruoss, A., Duquenne, P.-A., Catt, E., Genewein, T., Mattern, C., Grau-Moya, J., Wenliang, L.~K., Aitchison, M., Orseau, L., Hutter, M., and Veness, J.
\newblock Language modeling is compression.
\newblock In \emph{{Proc. of the 12th International Conference on Learning Representations (ICLR'24)}}, 2024.

\bibitem[Fiil-Flynn et~al.(2022)Fiil-Flynn, Butler, Carroll, Cohen-Sasson, Craig, Guibault, Jaszi, Jütte, Katz, Quintais, Margoni, de~Souza, Sag, Samberg, Schirru, Senftleben, Tur-Sinai, and Contreras]{fiilflynn2022legal}
Fiil-Flynn, S.~M., Butler, B., Carroll, M., Cohen-Sasson, O., Craig, C., Guibault, L., Jaszi, P., Jütte, B.~J., Katz, A., Quintais, J.~P., Margoni, T., de~Souza, A.~R., Sag, M., Samberg, R., Schirru, L., Senftleben, M., Tur-Sinai, O., and Contreras, J.~L.
\newblock Legal reform to enhance global text and data mining research.
\newblock \emph{Science}, 378\penalty0 (6623):\penalty0 951--953, 2022.

\bibitem[Franceschelli \& Musolesi(2022)Franceschelli and Musolesi]{franceschelli2022copyright}
Franceschelli, G. and Musolesi, M.
\newblock Copyright in generative deep learning.
\newblock \emph{Data \& Policy}, 4:\penalty0 e17, 2022.

\bibitem[{Gemini Team} et~al.(2023){Gemini Team}, Anil, Borgeaud, Wu, Alayrac, Yu, Soricut, Schalkwyk, Dai, Hauth, et~al.]{team2023gemini}
{Gemini Team}, Anil, R., Borgeaud, S., Wu, Y., Alayrac, J.-B., Yu, J., Soricut, R., Schalkwyk, J., Dai, A.~M., Hauth, A., et~al.
\newblock Gemini: a family of highly capable multimodal models.
\newblock \emph{arXiv preprint arXiv:2312.11805}, 2023.

\bibitem[Grimmelman(2016)]{grimmelman2016thing}
Grimmelman, J.
\newblock There’s no such thing as a computer-authored work–and it’s a good thing, too.
\newblock \emph{The Columbia Journal of Law \& the Arts}, 39\penalty0 (3):\penalty0 403–416, 2016.

\bibitem[Guadamuz(2017)]{guadamuz2017androids}
Guadamuz, A.
\newblock Do androids dream of electric copyright? comparative analysis of originality in artificial intelligence generated works.
\newblock \emph{Intellectual Property Quarterly}, 2:\penalty0 1--24, 2017.

\bibitem[Guadamuz(2024)]{guadamuz2024scanner}
Guadamuz, A.
\newblock A scanner darkly: Copyright liability and exceptions in artificial intelligence inputs and outputs.
\newblock \emph{GRUR International}, 73\penalty0 (2):\penalty0 111--127, 2024.

\bibitem[Henderson et~al.(2023)Henderson, Li, Jurafsky, Hashimoto, Lemley, and Liang]{henderson2023foundation}
Henderson, P., Li, X., Jurafsky, D., Hashimoto, T., Lemley, M.~A., and Liang, P.
\newblock Foundation models and fair use.
\newblock \emph{Journal of Machine Learning Research}, 24\penalty0 (400):\penalty0 1--79, 2023.

\bibitem[Lee et~al.(2024)Lee, Cooper, and Grimmelmann]{lee2024talkin}
Lee, K., Cooper, A.~F., and Grimmelmann, J.
\newblock Talkin' 'bout {AI} generation: Copyright and the generative-{AI} supply chain, 2024.
\newblock Forthcoming, \textit{Journal of the Copyright Society of the USA '24}.

\bibitem[Leike et~al.(2018)Leike, Krueger, Everitt, Martic, Maini, and Legg]{leike2018scalable}
Leike, J., Krueger, D., Everitt, T., Martic, M., Maini, V., and Legg, S.
\newblock Scalable agent alignment via reward modeling: a research direction, 2018.
\newblock arXiv:1811.07871 [cs.LG].

\bibitem[Lemley \& Casey(2021)Lemley and Casey]{lemley2021fair}
Lemley, M.~A. and Casey, B.
\newblock Fair learning.
\newblock \emph{Texas Law Review}, 99\penalty0 (4):\penalty0 743--785, 2021.

\bibitem[MacKay(2003)]{mackay2003information}
MacKay, D.~J.
\newblock \emph{Information Theory, Inference and Learning algorithms}.
\newblock Cambridge University Press, 2003.

\bibitem[{Meta}(2024)]{meta2024introducing}
{Meta}.
\newblock {Introducing Meta Llama 3: The most capable openly available LLM to date}, 2024.
\newblock \url{https://ai.meta.com/blog/meta-llama-3/} [accessed June 26, 2024].

\bibitem[Otero(2021)]{otero2021machine}
Otero, B.~G.
\newblock Machine {L}earning models under the {C}opyright microscope: Is {EU} {C}opyright fit for purpose?
\newblock \emph{GRUR International}, 70\penalty0 (11):\penalty0 1043--1055, 2021.

\bibitem[Rombach et~al.(2022)Rombach, Blattmann, Lorenz, Esser, and Ommer]{rombach2022high}
Rombach, R., Blattmann, A., Lorenz, D., Esser, P., and Ommer, B.
\newblock High-resolution image synthesis with latent diffusion models.
\newblock In \emph{Proc. of the IEEE/CVF Conference on Computer Vision and Pattern Recognition (CVPR'22)}, pp.\  10684--10695, 2022.

\bibitem[Samuelson(1986)]{samuelson1986allocating}
Samuelson, P.
\newblock Allocating ownership rights in computer-generated works.
\newblock In \emph{{Symposium on The Future of Software Protection}}, 1986.

\bibitem[Samuelson(2021)]{samuelson2021copyright}
Samuelson, P.
\newblock Text and data mining of in-copyright works: is it legal?
\newblock \emph{Communications of the ACM}, 64\penalty0 (11):\penalty0 20–22, 2021.

\bibitem[Saxe et~al.(2018)Saxe, Bansal, Dapello, Advani, Kolchinsky, Tracey, and Cox]{michael2018on}
Saxe, A.~M., Bansal, Y., Dapello, J., Advani, M., Kolchinsky, A., Tracey, B.~D., and Cox, D.~D.
\newblock On the information bottleneck theory of deep learning.
\newblock In \emph{{Proc. of the 6th International Conference on Learning Representations (ICLR'18)}}, 2018.

\bibitem[Schwarzschild et~al.(2024)Schwarzschild, Feng, Maini, Lipton, and Kolter]{schwarzschild2024rethinking}
Schwarzschild, A., Feng, Z., Maini, P., Lipton, Z., and Kolter, J.~Z.
\newblock Rethinking {LLM} memorization through the lens of adversarial compression.
\newblock In \emph{{Advances in Neural Information Processing Systems (NIPS'24)}}, 2024.

\bibitem[Shanahan(2024)]{shanahan2024talking}
Shanahan, M.
\newblock Talking about large language models.
\newblock \emph{Communications of the ACM}, 67\penalty0 (2):\penalty0 68–79, 2024.

\bibitem[Shumailov et~al.(2024)Shumailov, Shumaylov, Zhao, Papernot, Anderson, and Gal]{shumailov2024ai}
Shumailov, I., Shumaylov, Z., Zhao, Y., Papernot, N., Anderson, R., and Gal, Y.
\newblock {AI} models collapse when trained on recursively generated data.
\newblock \emph{Nature}, 631\penalty0 (8022):\penalty0 755--759, 2024.

\bibitem[Shwartz-Ziv \& LeCun(2024)Shwartz-Ziv and LeCun]{shwartzziv2024compress}
Shwartz-Ziv, R. and LeCun, Y.
\newblock To compress or not to compress—self-supervised learning and information theory: A review.
\newblock \emph{Entropy}, 26\penalty0 (3):\penalty0 252, 2024.

\bibitem[Shwartz-Ziv \& Tishby(2017)Shwartz-Ziv and Tishby]{shwartzziv2017opening}
Shwartz-Ziv, R. and Tishby, N.
\newblock Opening the black box of deep neural networks via information, 2017.
\newblock arXiv:1703.00810 [cs.LG].

\bibitem[Sobel(2021)]{sobel2020taxonomy}
Sobel, B.
\newblock A taxonomy of training data: Disentangling the mismatched rights, remedies, and rationales for restricting machine learning.
\newblock In \emph{Artificial Intelligence and Intellectual Property}, pp.\  221–242. Oxford University Press, 2021.

\bibitem[{Sousa e Silva}(2024)]{sousa2024ai}
{Sousa e Silva}, N.
\newblock Are {AI} models’ weights protected databases?, 2024.
\newblock \url{https://copyrightblog.kluweriplaw.com/2024/01/18/are-ai-models-weights-protected-databases/} [last access: May 27, 2024].

\bibitem[Tishby \& Zaslavsky(2015)Tishby and Zaslavsky]{tishby2015deep}
Tishby, N. and Zaslavsky, N.
\newblock Deep learning and the information bottleneck principle.
\newblock In \emph{{Proc. of the 2015 IEEE Information Theory Workshop (ITW'15)}}, 2015.

\bibitem[Tishby et~al.(1999)Tishby, Pereira, and Bialek]{tishby1999information}
Tishby, N., Pereira, F.~C., and Bialek, W.
\newblock The information bottleneck method.
\newblock In \emph{{Proc. of the 37th Annual Allerton Conference on Communication, Control and Computing}}, 1999.

\bibitem[Tunstall et~al.(2022)Tunstall, von Werra, and Wolf]{tunstall2022natural}
Tunstall, L., von Werra, L., and Wolf, T.
\newblock \emph{Natural Language Processing with Transformers}.
\newblock O'Reilly, 2022.

\bibitem[Vyas et~al.(2023)Vyas, Kakade, and Barak]{vyas2023provable}
Vyas, N., Kakade, S.~M., and Barak, B.
\newblock On provable copyright protection for generative models.
\newblock In \emph{{Proc. of the 40th International Conference on Machine Learning (ICML'23)}}, 2023.

\bibitem[Weidinger et~al.(2022)Weidinger, Uesato, Rauh, Griffin, Huang, Mellor, Glaese, Cheng, Balle, Kasirzadeh, Biles, Brown, Kenton, Hawkins, Stepleton, Birhane, Hendricks, Rimell, Isaac, Haas, Legassick, Irving, and Gabriel]{weidinger2022taxonomy}
Weidinger, L., Uesato, J., Rauh, M., Griffin, C., Huang, P.-S., Mellor, J., Glaese, A., Cheng, M., Balle, B., Kasirzadeh, A., Biles, C., Brown, S., Kenton, Z., Hawkins, W., Stepleton, T., Birhane, A., Hendricks, L.~A., Rimell, L., Isaac, W., Haas, J., Legassick, S., Irving, G., and Gabriel, I.
\newblock Taxonomy of risks posed by language models.
\newblock In \emph{{Proc. of the 2022 {ACM} Conference on Fairness, Accountability, and Transparency (FAccT'22})}, pp.\  214–229, 2022.

\bibitem[{World Trade Organization}(1994)]{wto1994agreement}
{World Trade Organization}.
\newblock {Agreement on Trade-Related Aspects of Intellectual Property Rights}, 1994.

\end{thebibliography}
\bibliographystyle{icml2024}

\end{document}